\documentclass[prd,aps,preprint,showpacs,nofootinbib,superscriptaddress]{revtex4-1}
\usepackage{epsfig}
\usepackage{amsmath, slashed}
\usepackage{color}

\newcommand{\be}[1]{
\begin{eqnarray}\label{#1}}
\newcommand{\ee}{\end{eqnarray}}

\allowdisplaybreaks
\begin{document}
%\preprint{xxx}\preprint{xxx}

\title{Twist-3 GPDs in Deeply-Virtual Compton Scattering}

\author{Fatma Aslan} 
\affiliation{Department of Physics, New Mexico State University,
                Las Cruces, NM 88003, USA}

\author{Matthias Burkardt} 
\affiliation{Department of Physics, New Mexico State University,
                Las Cruces, NM 88003, USA}

\author{C\'edric Lorc\'e}
\affiliation{Centre de Physique Th\'eorique, \'Ecole Polytechnique, CNRS, 
              Universit\'e Paris-Saclay, 91128 Palaiseau, France}

\author{Andreas Metz}
\affiliation{Department of Physics, SERC,
             Temple University, Philadelphia, PA 19122, USA}
             
\author{Barbara Pasquini}
\affiliation{Dipartimento di Fisica, Universit\`a degli Studi di Pavia,
               27100 Pavia, Italy}
\affiliation{Istituto Nazionale di Fisica Nucleare, Sezione di Pavia,
               27100 Pavia, Italy}

\begin{abstract}
The sub-leading power of the scattering amplitude for deeply-virtual Compton scattering (DVCS) off the nucleon contains leading-twist and twist-3 generalized parton distributions (GPDs).
We point out that in DVCS, at twist-3 accuracy, one cannot address any individual twist-3 GPD.
This complication appears on top of the deconvolution issues familiar from the twist-2 DVCS amplitude.
Accessible are exclusively linear combinations involving both vector and axial-vector twist-3 GPDs.
This implies, in particular, that the (kinetic) orbital angular momentum of quarks can hardly be constrained by twist-3 DVCS observables.
Moreover, using the quark-target model, we find that twist-3 GPDs can be discontinuous.
The discontinuities however cancel in the DVCS amplitude, which further supports the hypothesis of factorization at twist-3 accuracy.
\end{abstract}

%\pacs{12.15.-y; 12.38.-t; 12.39.St; 13.85.-t; 13.88.+e}

\date{\today}

\maketitle

%%%%%%%%%%%%%%%%%%%%
\section{Introduction} 
\label{s:intro}
%%%%%%%%%%%%%%%%%%%%
It has been known for more than two decades that deeply-virtual Compton scattering (DVCS) off the nucleon, i.e.~the process $\gamma^{\ast} N \to \gamma N$, opens up new avenues for exploring the parton structure of the nucleon~\cite{Mueller:1998fv, Ji:1996ek, Radyushkin:1996nd, Ji:1996nm}. 
It was found that the scattering amplitude of DVCS can be expressed in terms of generalized parton distributions (GPDs)~\cite{Mueller:1998fv, Ji:1996ek, Radyushkin:1996nd, Ji:1996nm, Radyushkin:1996ru, Collins:1998be}, a novel type of functions which not only contain all the physics encoded in ordinary parton distributions and in form factors but also genuine new information --- see Refs.~\cite{Goeke:2001tz, Diehl:2003ny, Belitsky:2005qn, Boffi:2007yc, Guidal:2013rya, Mueller:2014hsa, Kumericki:2016ehc} for reviews on GPDs.
In particular, through leading-twist (twist-2) GPDs one can access the angular momentum of quarks and gluons inside hadrons~\cite{Ji:1996ek}, and explore the 3-dimensional parton structure of hadrons~\cite{Burkardt:2000za,Ralston:2001xs,Diehl:2002he,Burkardt:2002hr}.
\\
\indent
In order to extract twist-2 GPDs from data on DVCS one must have sufficient control over power corrections to the leading-twist amplitude.
This applies the more so if the (negative) squared four-momentum of the virtual photon is not very large, as is often the case in past and scheduled experiments --- see for instance Refs.~\cite{Airapetian:2001yk, Stepanyan:2001sm, Camacho:2006qlk, Airapetian:2012pg, Jo:2015ema, Defurne:2015kxq, Akhunzyanov:2018nut}.
Quite some effort has therefore been devoted to get a detailed understanding of effects in DVCS that appear at twist-3 level and beyond~\cite{Anikin:2000em, Penttinen:2000dg, Belitsky:2000vx, Kivel:2000rb, Radyushkin:2000jy, Kivel:2000cn, Radyushkin:2000ap, Belitsky:2000vk, Kivel:2000fg, White:2001pu, Anikin:2001ge, Belitsky:2001yp, Radyushkin:2001fc, Kivel:2001rw, Geyer:2001qf, Belitsky:2001ns, Kiptily:2002nx, Kivel:2003jt, Freund:2003qs, Geyer:2004bx, Anikin:2009hk, Belitsky:2010jw, Braun:2012bg, Braun:2012hq, Braun:2014sta}.
\\
\indent
Power corrections to the leading-twist DVCS amplitude also contain genuine new information about the hadron structure.
In fact, at twist-3 level in DVCS off the nucleon eight twist-3 GPDs show up~\cite{Anikin:2000em, Penttinen:2000dg, Belitsky:2000vx, Kivel:2000cn, Kivel:2000fg}. 
So far four major motivations for measuring twist-3 GPDs have been put forward in the literature.
First, there is a relation between one particular twist-3 GPD and the orbital angular momentum (OAM) of quarks inside a longitudinally polarized nucleon~\cite{Penttinen:2000dg}. 
In the notation of Ref.~\cite{Kiptily:2002nx} one has (for each quark flavor $q$)
\begin{align}
L_{\rm kin}^q = - \int_{-1}^{1} d x \, x \, G_2^q(x, \xi = 0, t = 0) \,,
\label{e:OAM_relation}
\end{align}
where the twist-3 GPD $G_2$ depends on the (average) longitudinal quark momentum $x$, as well as the longitudinal ($\xi$) and total ($t$) momentum transfer to the nucleon.
Note that in Eq.~\eqref{e:OAM_relation} enters the so-called kinetic OAM $L_{\rm kin}^q$ as defined by Ji in Ref.~\cite{Ji:1996ek}, which is to be distinguished from the canonical OAM $L_{\rm can}^q$ of Jaffe and Manohar~\cite{Jaffe:1989jz}.
More information on the spin decomposition of the nucleon can be found in recent review articles~\cite{Leader:2013jra, Wakamatsu:2014zza, Liu:2015xha} and in Ref.~\cite{Burkardt:2012sd}, where a physical interpretation of the difference between $L_{\rm kin}^q$ and $L_{\rm can}^q$ was given.
According to~\cite{Hatta:2012cs}, $L_{\rm can}^q$ can also be related to twist-3 off-forward matrix elements that are defined through quark-gluon-quark operators.  
But here we concentrate on $L_{\rm kin}^q$ and the GPD $G_2$ which appears in the parameterization of the off-forward quark-quark correlator and was shown to enter the twist-3 amplitude of DVCS.
The relation in~\eqref{e:OAM_relation} can be considered an alternative to Ji's relation between $L_{\rm kin}^q$ and twist-2 GPDs~\cite{Ji:1996ek}.
At the very least it could be used for cross-checks. 
Another motivation for exploring twist-3 GPDs is a relation to the (average) transverse force acting on a quark in a polarized nucleon~\cite{Burkardt:2008ps, Burkardt:prep}.
Third, certain spin-orbit correlations in the nucleon can be expressed through twist-3 GPDs~\cite{Lorce:2014mxa, Lorce:2011kd, Bhoonah:2017olu}.
Fourth, in Refs.~\cite{Rajan:2016tlg, Rajan:2017xlo} some relations have been obtained between twist-3 GPDs and generalized transverse momentum dependent parton distributions (GTMDs)~\cite{Meissner:2008ay, Meissner:2009ww, Lorce:2013pza}, which in principle allow one to constrain the latter functions through the former.
 \\
 \indent
In this work we show that one cannot address any individual twist-3 GPD via the DVCS process.
(Note that here we focus exclusively on GPDs of quarks.  
Gluon GPDs, which enter the DVCS amplitude at higher order in the strong coupling, can also play an important role for the DVCS phenomenology, especially at higher energies --- see for instance Ref.~\cite{Moutarde:2013qs}.)
Irrespective of the parameterization of the GPDs and the (polarization) observable under discussion, in DVCS one can exclusively access linear combinations that involve both vector and axial-vector twist-3 GPDs.
This implies, in particular, that in DVCS one cannot measure $L_{\rm kin}^q$ through the twist-3 GPD $G_2$, which is in contrast to some hopes/claims expressed in the literature --- see for instance Refs.~\cite{Penttinen:2000dg, Kivel:2000cn, Courtoy:2013oaa, Courtoy:2014bea, Pisano:2016dnz}. 
One might therefore resort to other processes, such as double DVCS, where discontinuities in the GPDs do not appear to cause any problem~\cite{Kivel:2000rb}.
\\
\indent
Irrespective of whether individual twist-3 GPDs can be measured, it is important to explore QCD factorization of the DVCS amplitude at twist-3 accuracy.
The leading-order (LO) expression of the DVCS amplitude, {\it a priori}, only provides limited insight in that regard.
Nevertheless, the LO result already shows that factorization is endangered if the GPDs are discontinuous at $x = \pm \, \xi$.
Various studies have used the Wandzura-Wilczek (WW) approximation~\cite{Wandzura:1977qf} for twist-3 GPDs and the twist-3 DVCS amplitude~\cite{Belitsky:2000vx, Kivel:2000rb, Radyushkin:2000jy, Kivel:2000cn, Radyushkin:2000ap, Belitsky:2000vk, Kivel:2000fg, Anikin:2001ge, Belitsky:2001yp, Kivel:2001rw, Belitsky:2001ns, Kiptily:2002nx, Kivel:2003jt, Freund:2003qs}.
The WW approximation does actually lead to discontinuous twist-3 GPDs~\cite{Kivel:2000rb, Radyushkin:2000jy, Kivel:2000cn, Radyushkin:2000ap}, as we make explicit below for all 
twist-3 GPDs of the nucleon\footnote{
In Refs.~\cite{Theussl:2002xp, Broniowski:2007si} a discontinuous result for the twist-2 GPD $H$ of the pion was found in the Nambu--Jona-Lasinio model.
This model calculation is the only one we are aware of leading to a discontinuous twist-2 GPD.}.
However, the discontinuities cancel between different terms in the DVCS amplitude~\cite{Kivel:2000rb, Radyushkin:2000jy, Kivel:2000cn, Radyushkin:2000ap} so that one has factorization at LO. 
In Ref.~\cite{Kivel:2003jt}, part of the NLO amplitude for DVCS off the nucleon was computed in the WW approximation and found to factorize as well.
\\
\indent
One may wonder if discontinuities of twist-3 GPDs are an artifact of the WW approximation.
However, we show that also in the quark-target model (QTM) twist-3 GPDs are discontinuous.
This result again brings up the question about factorization of the DVCS amplitude at twist-3 accuracy.
But, like in the case of the WW approximation, the linear combinations of GPDs that enter the DVCS amplitude are well-behaved.
This finding supports the hypothesis of factorization of the twist-3 DVCS amplitude.
On the other hand, it also shows that a  phenomenological study of twist-3 DVCS observables where individual GPDs are varied independently is
not practicable, because of the delicate cancellation of discontinuities which occurs in the linear combinations of twist-3 GPDs.
\\
\indent
The remainder of the paper is organized as follows: 
In Sec.~\ref{s:amplitude}, we recall the Compton tensor for DVCS at twist-3 accuracy and derive the linear combinations of twist-3 GPDs that can be addressed.
In Sec.~\ref{s:WWapproximation}, we give a brief discussion about twist-3 GPDs in the WW approximation, while Sec.~\ref{s:QTM} contains our results for twist-3 GPDs in the QTM.
We summarize the work in Sec.~\ref{s:summary}.
Relations between certain Dirac bilinears and between different parameterizations of twist-3 GPDs, as well as more details about the WW approximation can be found in the appendices.

%%%%%%%%%%%%%%%%%%%%
\section{DVCS amplitude of the nucleon at twist-3 accuracy}
\label{s:amplitude}
%%%%%%%%%%%%%%%%%%%%
We now discuss the amplitude of virtual Compton scattering off the nucleon,
\begin{equation}
\gamma^*(q) + N(p) \to \gamma(q') + N(p') \, ,
\label{e:process}
\end{equation}
where the four-momenta of the particles are indicated, while spin labels are omitted for brevity.
One has $p^2 = p'^2 = m^2$, with $m$ denoting the nucleon mass, and $t = (p - p')^2$.
We are considering a reference frame in which the average nucleon momentum $P = \frac{1}{2}(p + p')$ and the momentum of the virtual photon have no transverse components.
This allows one to write~\cite{Goeke:2001tz, Kivel:2003jt} 
\begin{equation}
P = n^* + \frac{\bar{m}^2}{2} \, n \,, \qquad
q = - 2 \, \xi' n^* + \frac{Q^2}{4 \xi'} \, n \,, \qquad
\Delta = p' - p = - 2 \, \xi n^* + \xi \bar{m}^2 \, n + \Delta_\perp \,,
\label{e:kinem}
\end{equation}
with $Q^2 = - q^2$.
According to~\eqref{e:kinem}, the four-momenta $P$ and $q$ specify two light-like vectors ($n$, $n^*$) which satisfy
\begin{equation}
n \cdot n = 0\,, \qquad 
n^* \cdot n^* = 0 \,, \qquad
n \cdot n^* = 1 \,.
\label{e:lightlike}
\end{equation}
This also implies $P^2 = \bar{m}^2 = m^2 - \frac{t}{4}$.
We define the transverse metric tensor and anti-symmetric epsilon tensor through
\begin{equation}
g^{\mu \nu}_\perp = g^{\mu \nu} - n^\mu \, n^{* \nu} - n^\nu \, n^{* \mu} \,, \qquad
\varepsilon_\perp^{\mu \nu} = \varepsilon^{\mu \nu \alpha \beta} \, n_\alpha \, n^*_\beta \, ,
\label{e:metric}
\end{equation}
where $\varepsilon_{\mu \nu \alpha\beta}$ is the totally antisymmetric Levi-Civita tensor.
(We use $\varepsilon_{0123} = +1$.)
By means of $g_\perp^{\mu\nu}$ one can introduce transverse four-vectors\footnote{The light-cone plus-momentum and minus-momentum of an arbitrary four-vector $v$ are defined according to $v^+ = \frac{1}{\sqrt{2}} (v^0 + v^3) = P^+ \, n \cdot v $ and $v^- = \frac{1}{\sqrt{2}} (v^0 - v^3) = \frac{1}{P^+} \, n^* \cdot v$, respectively.}.
In particular, the four-vector of the transverse momentum transfer of the nucleon in Eq.~\eqref{e:kinem} is given by $\Delta_\perp^\mu = g_\perp^{\mu\nu} \Delta_\nu$, with $\Delta_\perp^2 = - \vec{\Delta}_\perp^2$.
For the variables $\xi'$ and $\xi$ in~\eqref{e:kinem} one has
\begin{equation}
\xi' = \frac{x_B}{2 - x_B} + {\cal O}(1 / Q^2) \,, \qquad 
\xi' = \xi + {\cal O}(1 / Q^2) \,,
\label{e:kinem_approx}
\end{equation}
where $x_B = Q^2 / (2 p \cdot q)$.
The exact expressions for the correction terms in~\eqref{e:kinem_approx} can be found in Refs.~\cite{Goeke:2001tz, Kivel:2003jt}.
We also note that to twist-3 accuracy one can use
\begin{equation}
P = n^* \,, \qquad
q = - 2 \, \xi P + \frac{Q^2}{4 \xi} \, n \,, \qquad
\Delta = p' - p = - 2 \, \xi P + \Delta_\perp \,,
\label{e:kinem_twist3}
\end{equation}
instead of the equations in~\eqref{e:kinem}~\cite{Kivel:2000cn, Kivel:2000fg}.
\\
\indent
The scattering amplitude for DVCS follows from the Compton tensor $T^{\mu \nu}$, which in turn is defined through the matrix element of the time-ordered product of two electromagnetic currents,
\begin{equation}
T^{\mu\nu} = - \, i \int d^4x\ e^{-i \, q \cdot x}
\langle p' | \, T\left[J_{\rm e.m.}^\mu (x) J_{\rm e.m.}^\nu(0)\right] | p\rangle \, ,
\label{e:T_def}
\end{equation}
where the index $\mu$ ($\nu$) refers to the virtual (real) photon.
This Compton tensor at twist-3 accuracy has been studied by several groups using different methods~\cite{Anikin:2000em, Penttinen:2000dg, Belitsky:2000vx, Radyushkin:2000jy, Kivel:2000cn, Kivel:2000fg, Belitsky:2001yp}.
In the generalized Bjorken limit $Q^2 \to \infty$, $2 \, p\cdot q \to \infty$, with $x_B$ constant, and $|t| \ll Q^2$, the tensor $T^{\mu\nu}$ of the nucleon, through ${\cal O}(1/Q)$ accuracy, takes the form~\cite{Kivel:2000cn, Kivel:2000fg}\footnote{In~Eq.~\eqref{e:T_res}, we have omitted flavor labels for $F^\mu$ and $\widetilde{F}^\mu$, and the overall sum $\sum_q e_q^2$ where $e_q$ is the quark charge in units of the elementary charge.}
\begin{eqnarray}
T^{ \mu\nu} & = & 
\frac{1}{2} \int_{-1}^1 dx \; \bigg[ 
\bigg( \!\! - g^{\mu\nu}_\perp - \frac{P^\nu \Delta_\perp^\mu}{P \cdot q'}  \bigg) n^\beta F_\beta (x, \xi, \Delta) \, C^+(x, \xi) 
\nonumber \\[0.2cm]
& + & \bigg( \!\! -g_\perp^{\nu\alpha} - \frac{P^\nu \Delta_\perp^\alpha}{P \cdot q'} \bigg)
i \varepsilon_{\perp \alpha}^{\mu} \, n^\beta \widetilde{F}_\beta (x, \xi, \Delta) \, C^-(x, \xi) 
\nonumber \\[0.2cm]
& - & \frac{(q + 4\xi P)^\mu}{P \cdot q} \bigg( \!\! - g_\perp^{\nu\alpha} - \frac{P^\nu \Delta_\perp^\alpha}{P \cdot q'}  \bigg) \!
\Big( F_\alpha (x, \xi, \Delta) \, C^+(x, \xi) -  i \varepsilon_{\perp\alpha\beta} \widetilde{F}^\beta (x, \xi, \Delta) \, C^-(x, \xi) \! \Big) \bigg] \,, \phantom{aa}
\label{e:T_res}
\end{eqnarray}
with the matrix elements $F^\mu$ and $\widetilde{F}^\mu$, for a quark flavor $q$, defined as
\begin{eqnarray} 
F_q^\mu (x, \xi, \Delta) & = & \int_{-\infty}^\infty \frac{d \lambda}{2\pi} \, e^{-i\lambda x} \,
\langle p' | \, \bar{q}(\tfrac{\lambda}{2} n) \, \gamma^\mu \, {\cal W}(\tfrac{\lambda}{2}n, - \tfrac{\lambda}{2}n) \, q(-\tfrac{\lambda}{2} n) \, | p \rangle \,,
\label{e:vector}
\\[0.2cm] 
\widetilde{F}_q^\mu (x, \xi, \Delta) & = & \int_{-\infty}^\infty \frac{d \lambda}{2\pi} \, e^{-i\lambda x} \,
\langle p' | \, \bar{q}(\tfrac{\lambda}{2} n) \, \gamma^\mu \gamma_5 \, {\cal W}(\tfrac{\lambda}{2}n, - \tfrac{\lambda}{2}n) \, q(-\tfrac{\lambda}{2} n) \, | p \rangle \,.
\label{e:axial}
\end{eqnarray}
In Eqs.~\eqref{e:vector} and~\eqref{e:axial}, ${\cal W}(\tfrac{\lambda}{2}n, - \tfrac{\lambda}{2}n)$ indicates a straight Wilson line which ensures gauge invariance of the operators.
The (LO) coefficient functions in~\eqref{e:T_res} are
\begin{equation}
C^\pm(x,\xi) = \frac{1}{x - \xi + i \varepsilon} \pm \frac{1}{x + \xi - i \varepsilon} \,.
\label{e:coeff}
\end{equation}
The expression in Eq.~\eqref{e:T_res} agrees with the result in Refs.~\cite{Belitsky:2000vx, Belitsky:2001yp}.   
\\
\indent
Up to and including twist-3 effects, the correlators in Eq.~\eqref{e:vector} and Eq.~\eqref{e:axial} can be decomposed into six vector GPDs and six axial-vector GPDs, respectively. 
Using the definition of GPDs from Ref.~\cite{Kiptily:2002nx}, one has 
\begin{eqnarray}
F^\mu & = & P^\mu \frac{h^+}{P^+} \, H + P^\mu \frac{e^+}{P^+} \,E
\nonumber \\[0.2cm]
& & + \, \Delta_\perp^\mu \frac{b}{2m} \, G_1 + h_\perp^\mu \, (H + E + G_2) + \Delta_\perp^\mu \frac{h^+}{P^+} \, G_3
+ \tilde{\Delta}_\perp^\mu \frac{\tilde h^+}{P^+} \, G_4 \,, 
\label{e:tw3_vector}
\\[0.2cm]
\widetilde{F}^\mu & = & P^\mu \frac{\tilde h^+}{P^+} \, \widetilde{H} + P^\mu \frac{\tilde e^+}{P^+} \, \widetilde{E}
\nonumber \\[0.2cm]
& & + \, \Delta_\perp^\mu \frac{\tilde{b}}{2m} \, (\widetilde{E} + \widetilde{G}_1) + \tilde h_\perp^\mu \, (\widetilde{H} + \widetilde{G}_2) 
+ \Delta^\mu_\perp \frac{\tilde{h}^+}{P^+} \, \widetilde{G}_3
+ \tilde{\Delta}_\perp^\mu \frac{h^+}{P^+} \, \widetilde{G}_4 \,,
\label{e:tw3_axial}
\end{eqnarray}
where, as is well known, the GPDs $H$, $E$ ($\widetilde{H}$, $\widetilde{E}$) fully specify the leading-twist contribution of the correlator $F^\mu$ ($\widetilde{F}^\mu$).
The (new) vector GPDs $G_1, \ldots , G_4$ and axial-vector GPDs $\widetilde{G}_1, \ldots , \widetilde{G}_4$ enter at twist-3 accuracy.
In Eqs.~\eqref{e:tw3_vector},~\eqref{e:tw3_axial}, we have omitted the arguments of $F^\mu$, $\widetilde{F}^\mu$ and the GPDs, and we made use of the Dirac bilinears~\cite{Belitsky:2000vx}
\begin{equation}
\begin{aligned}
h^\mu &= \bar u(p') \, \gamma^\mu \, u ( p ) \,, \qquad & 
e^\mu &= \bar u(p') \, \frac{i \sigma^{\mu\nu} \Delta_\nu}{2m} \, u ( p ) \,,  \quad &
b &= \bar u(p') \, u ( p ) \,, 
\\[0.2cm]
\tilde h^\mu &= \bar u(p') \, \gamma^\mu \gamma_5 \, u ( p ) \,, &
\tilde e^\mu &= \frac{\Delta^\mu}{2m} \, \tilde{b} \,, &
\tilde b &= \bar u(p') \, \gamma_5 \, u ( p ) \,,
\end{aligned}
\label{e:bilinears}
\end{equation}
and the (transverse) four-vector $\tilde \Delta_\perp^\mu=i \varepsilon_{\perp}^{\mu\nu}\Delta_\nu$.
(For later convenience we also introduce $t^{\mu \nu} = \bar{u}(p') \, i \sigma^{\mu \nu} \, u(p)$, with $\sigma^{\mu \nu} = \frac{i}{2} [\gamma^\mu, \gamma^\nu]$. 
Relations between different Dirac bilinears are summarized in App.~A.)
Alternative definitions of twist-3 GPDs were introduced in Refs.~\cite{Belitsky:2001yp, Belitsky:2001ns} and in Ref.~\cite{Meissner:2009ww}.
In App.~B we give relations between the different sets of twist-3 GPDs.
\\
\indent
Twist-3 GPDs enter in the 3rd term of the r.h.s.~in Eq.~\eqref{e:T_res} only.
This term is suppressed for transversely polarized virtual photons~\cite{Kivel:2000rb, Radyushkin:2000jy, Kivel:2000cn}.
The DVCS amplitude at twist-3 accuracy therefore contains twist-3 GPDs for longitudinally polarized virtual photons only, while for transverse photon polarization the amplitude is entirely determined by twist-2 GPDs.
With the longitudinal polarization vector~\cite{Kivel:2000fg}
\begin{equation}
\varepsilon^\mu_L=\frac{1}{Q}(2\xi P^\mu+\frac{Q^2}{4\xi}n^\mu) \,,
\label{e:long_pol}
\end{equation}
one readily finds
\begin{eqnarray}
\varepsilon_{L \mu} \, T^{\mu\nu} 
& = & \frac{2\xi}{Q}
\int_{-1}^1 dx \, \big[ F^\nu_\perp \, C^+(x,\xi) - i \varepsilon_{\perp \alpha}^\nu \, \widetilde{F}_\perp^\alpha \, C^-(x,\xi) \big] 
\nonumber \\[0.2cm]
& = & \frac{2\xi}{Q}
\int_{-1}^1 dx \, \bigg[ \big( F^\nu_\perp - i \varepsilon_{\perp \alpha}^\nu \, \widetilde{F}_\perp^\alpha)  \, \frac{1}{x - \xi + i \varepsilon} 
+ \big( F^\nu_\perp + i \varepsilon_{\perp \alpha}^\nu \, \widetilde{F}_\perp^\alpha) \, \frac{1}{x + \xi - i \varepsilon} \bigg] \,,
\label{e:long_comb}
\end{eqnarray}
where we have neglected a power-suppressed term.
(In the next section we will make use of the last line in~\eqref{e:long_comb}.) 
Since our main interest is in the contribution of twist-3 GPDs to the DVCS amplitude we focus in the following on the expression in Eq.~\eqref{e:long_comb}.
Using the parametrizations in Eqs.~\eqref{e:tw3_vector} and~\eqref{e:tw3_axial}, and the relations in~\eqref{e:dirac_1a} and~\eqref{e:dirac_2a} one 
can write the integral in~\eqref{e:long_comb} as
\begin{eqnarray}
\lefteqn{\int_{-1}^1 dx \, \big[ F^\nu_\perp \, C^+(x,\xi) - i \varepsilon_{\perp\alpha}^\nu \, \widetilde{F}_\perp^\alpha \, C^-(x,\xi) \big] }
\nonumber \\
& = & \int_{-1}^1 dx \, \bigg[
\Delta_\perp^\nu \frac{b}{2 m} \Big( G_1 \, C^+ + (\widetilde{E} + \widetilde{G}_1) \, C^- \Big)
\nonumber\\
& + & h_\perp^\nu \bigg( (H + E + G_2) \, C^+ - \frac{\Delta_\perp^2}{4 \xi m^2} \, (\widetilde{E} + \widetilde{G}_1) \, C^- - \frac{1}{\xi} \, (\widetilde{H} + \widetilde{G}_2) \, C^-  \bigg)
\nonumber\\
& + & \Delta_\perp^\nu \frac{h^+}{P^+} \bigg( G_3 \, C^+ - \frac{\bar{m}^2}{2 m^2} \, (\widetilde{E} + \widetilde{G}_1) \, C^- - \widetilde{G}_4 \, C^- \bigg)
\nonumber\\
& + & \tilde{\Delta}_\perp^\nu \frac{\tilde{h}^+}{P^+} \bigg( G_4 \, C^+ + \frac{t}{8 \xi m^2} \, (\widetilde{E} + \widetilde{G}_1) \, C^-  + \frac{1}{2\xi} \,(\widetilde{H} + \widetilde{G}_2) \, C^- - \widetilde{G}_3 \, C^- \bigg) \bigg] \,.
\label{e:long_comb_res}
\end{eqnarray}
Equation~\eqref{e:long_comb_res} shows explicitly that, at twist-3 accuracy, twist-3 GPDs enter through four independent structures only.
This result is obviously independent of the polarization of the particles in the DVCS process.
In particular, one always has linear combinations of both vector and axial-vector twist-3 GPDs.
Using different parameterizations of the GPDs will therefore not alter the situation.
While this general finding is implied by previous work~\cite{Penttinen:2000dg, Belitsky:2000vx, Kivel:2000cn, Belitsky:2001yp}, to the best of our knowledge it has never been made explicit through an equation of the type~\eqref{e:long_comb_res} that no individual twist-3 GPD can be measured directly through the DVCS process.
\\
\indent
It is well known that the leading-twist Compton tensor also contains both vector and axial-vector (twist-2) GPDs.
In that case, however, the GPDs can be disentangled because the two types of GPDs are associated with two independent tensors --- $g_\perp^{\mu\nu}$ for the vector GPDs, and $\varepsilon_\perp^{\mu\nu}$ for the axial-vector GPDs.
\\
\indent
Our finding affects all the motivations for studying twist-3 GPDs mentioned in the Introduction.
In particular, in DVCS at twist-3 accuracy there is no direct access to the kinetic OAM $L_{\rm kin}^q$ through the GPD $G_2$.
Specifically, in order to isolate $G_2$ one would need input not only for twist-2 GPDs but, according to Eq.~\eqref{e:long_comb_res}, also for the twist-3 GPDs $\widetilde{G}_1$ and $\widetilde{G}_2$.
This further complicates attempts to obtain information on $G_2$ from DVCS data.
In fact, as we argue below, since twist-3 GPDs can be discontinuous at $x = \pm \, \xi$, the situation is even more difficult.

%%%%%%%%%%%%%%%%%%%%
\section{Wandzura-Wilczek approximation}
\label{s:WWapproximation}
%%%%%%%%%%%%%%%%%%%%
A series of papers has studied the twist-3 DVCS amplitude in the WW approximation~\cite{Belitsky:2000vx, Kivel:2000rb, Radyushkin:2000jy, Kivel:2000cn, Radyushkin:2000ap, Belitsky:2000vk, Kivel:2000fg, Anikin:2001ge, Belitsky:2001yp, Kivel:2001rw, Belitsky:2001ns, Kiptily:2002nx, Kivel:2003jt, Freund:2003qs}.
In that approximation, twist-3 GPDs are decomposed into the so-called WW term, which is entirely given by twist-2 GPDs, and a contribution containing information about 3-parton (quark-gluon-quark) correlations in the nucleon.
Most of the equations for the WW approximation are summarized in App.~C, where we also list for the first time the WW term for all twist-3 vector and axial-vector GPDs of the nucleon.
\\
\indent
In the WW approximation, the twist-3 GPDs contain integrals that involve the WW kernels $W_{\pm}$ in~\eqref{e:Wpm} --- see Eqs.~\eqref{e:g1_ww}--\eqref{e:tg4_ww}.
These integrals generate discontinuities of the GPD correlators at $x = \pm \, \xi$, as was discussed for a spin-0 target in Refs.~\cite{Kivel:2000rb, Radyushkin:2000jy, Radyushkin:2000ap} and for a spin-$\frac{1}{2}$-target in Ref.~\cite{Kivel:2000cn}. 
To illustrate this point we consider the convolution
\begin{eqnarray}
f_{W_\pm}(x,\xi) & = & \int_{-1}^1 du \, W_\pm(x, u, \xi) \, f(u, \xi)
\nonumber \\[0.2cm]
& = & \frac{1}{2} \bigg[ \theta (x > \xi) \int_x^1 du \, \frac{f(u, \xi)}{u - \xi}  - \theta (x < \xi) \int_{-1}^x du \, \frac{f(u, \xi)}{u - \xi} \bigg]
\nonumber \\[0.2cm]
& & \pm \, \frac{1}{2} \bigg[ \theta (x > - \xi) \int_x^1 du \, \frac{f(u, \xi)}{u + \xi}  - \theta (x < - \xi) \int_{-1}^x du \, \frac{f(u, \xi)}{u + \xi} \bigg] \,,
\label{e:WW_conv}
\end{eqnarray}
with a generic function $f(u, \xi)$. 
Based on~\eqref{e:WW_conv} one readily derives
\begin{eqnarray}
\lim_{\delta \to 0} \Big[ f_{W_\pm}(\xi + \delta, \xi) - f_{W_\pm}(\xi - \delta,\xi) \Big]
& = & \frac{1}{2}  \, {\rm PV} \int_{-1}^1 du \, \frac{f(u, \xi)}{u - \xi} \,,
\label{e:plusxi_disc}
\\[0.2cm]
\lim_{\delta \to 0} \Big[ f_{W_\pm}(- \xi + \delta, \xi) - f_{W_\pm}(- \xi - \delta,\xi) \Big]
& = & \pm \, \frac{1}{2} \, {\rm PV} \int_{-1}^1 du \, \frac{f(u, \xi)}{u + \xi} \,.
\label{e:minusxi_disc}
\end{eqnarray}
Since the principal-value (PV) integrals on the r.h.s.~of~\eqref{e:plusxi_disc} and~\eqref{e:minusxi_disc} are generally nonzero, the quantities $f_{W_\pm}(x,\xi)$ are discontinuous at both $x = + \, \xi$ and $x = - \, \xi$.
Using the explicit expressions in Eqs.~\eqref{e:g1_ww}--\eqref{e:tg4_ww}, we therefore find that in the WW approximation all twist-3 vector and axial-vector GPDs of the nucleon have a discontinuity at $x = + \, \xi$ and at $x = - \, \xi$.
\\
\indent
The discontinuities of twist-3 GPDs endanger factorization of the DVCS amplitude in the WW approximation because integrals of the type
\begin{equation}
\int_{-1}^1 dx \, f_{W_\pm}(x, \xi) \, C^{\pm} (x, \xi) \,,
\label{e:undef_int}
\end{equation}
which appear in Eq.~\eqref{e:long_comb_res}, are obviously not defined~\cite{Kivel:2000rb}.
However, using the expressions in Eqs.~\eqref{e:vector_ww} and~\eqref{e:axial_ww}, plus the general results for the discontinuities in~\eqref{e:plusxi_disc} and~\eqref{e:minusxi_disc}, one finds that the linear combination 
$(F^\mu_\perp - i \varepsilon_{\perp \alpha}^\mu \, \widetilde{F}_\perp^\alpha)$ is continuous at $x = + \, \xi$, while
$(F^\mu_\perp + i \varepsilon_{\perp \alpha}^\mu \, \widetilde{F}_\perp^\alpha)$ is continuous at $x = - \, \xi$~\cite{Kivel:2000cn}. 
By means of the last line in Eq.~\eqref{e:long_comb} one then immediately verifies that the twist-3 DVCS amplitude of the nucleon is actually well-defined in the WW approximation~\cite{Kivel:2000cn}.
\\
\indent
The cancellation of discontinuities can of course also be discussed for the result in Eq.~\eqref{e:long_comb_res}, by using the WW term of the GPDs in Eqs.~\eqref{e:g1_ww}--\eqref{e:tg4_ww}.
For instance, in the first term on the r.h.s.~of Eq.~\eqref{e:long_comb_res} the two twist-3 GPDs $G_1$ and $\tilde{G}_1$ show up.
Based on the results in~\eqref{e:g1_ww} and~\eqref{e:tg1_ww} one finds\footnote{For the discussion of the WW approximation we have omitted the $t$-dependence of the GPDs.}
\begin{eqnarray} 
\lefteqn{\int_{-1}^{1} dx  \, \Big( G_1^{{\rm WW}} \, C^+ + (\widetilde{E} + \widetilde{G}_1^{{\rm WW}}) \, C^{-} \Big)}
\nonumber \\[0.2cm]
& = & \frac{1}{\xi} \int_{-1}^{1} dx \, \Big[ E(x,\xi) \, C^+(x,\xi)+\xi \widetilde{E}(x,\xi) \, C^-(x,\xi) \Big] 
\nonumber \\[0.2cm]
& + & \frac{1}{\xi} \int_{-1}^{1} dx \int_{-1}^{1} du \, \Big[ W_{+}(x,u,\xi) \, C^+(x,\xi) - W_{-}(x,u,\xi) \, C^-(x,\xi) \Big] \, {\cal D}_{u,\xi} \big[ E(u,\xi) \big] 
\nonumber \\[0.2cm]
& - & \frac{1}{\xi} \int_{-1}^{1} dx \int_{-1}^{1} du \, \Big[ W_{-}(x,u,\xi) \, C^+(x,\xi) - W_{+}(x,u,\xi) \, C^-(x,\xi) \, \Big] \, {\cal D}_{u,\xi} \big[ \xi \widetilde{E} (u,\xi) \big] \,.
\label{e:conv1}
\end{eqnarray}
Because of the linear combination of twist-3 GPDs, in Eq.~\eqref{e:conv1} one has two combinations of Wilson coefficients with the WW kernels only,
\begin{eqnarray} 
\lefteqn{\int _{-1}^{1} dx \int_{-1}^1 du \, \Big[ W_{\pm}(x,u,\xi) \, C^+(x,\xi) - W_{\mp}(x,u,\xi) \, C^-(x,\xi) \Big] \, f(u,\xi) }
\nonumber \\[0.2cm]
& = & \pm \int _{-1}^{1} dx \, \frac{1}{x - \xi + i \varepsilon} \, \int_{-1}^1 du \, \Big[ W_{+}(x,u,\xi) - W_{-}(x,u,\xi) \Big] \, f(u,\xi)
\nonumber \\[0.2cm]
& + & \int _{-1}^{1} dx \, \frac{1}{x + \xi - i \varepsilon} \, \int_{-1}^1 du \, \Big[ W_{+}(x,u,\xi) + W_{-}(x,u,\xi) \Big] \, f(u,\xi) \,.
\label{e:conv_WC}
\end{eqnarray}
The integrations upon $x$ in~\eqref{e:conv_WC} are well defined since, due to Eqs.~\eqref{e:plusxi_disc} and~\eqref{e:minusxi_disc}, the integrand of the first term on the r.h.s.~is continuous at $x = + \, \xi$, and the one of the second term is continuous at $x = - \, \xi$.
The exact same discussion applies to the other three terms on the r.h.s.~of Eq.~\eqref{e:long_comb_res}, as can be seen from the expressions in Eqs.~\eqref{e:conv2}--\eqref{e:conv4}.
The fact that in the WW approximation Eq.~\eqref{e:long_comb_res} is well-defined can be considered a consistency check of that equation and of the results in~\eqref{e:g1_ww}--\eqref{e:tg4_ww}.
\\
\indent
We emphasize that each twist-3 GPD in the four linear combinations that appear on the r.h.s.~of Eq.~\eqref{e:long_comb_res} is needed in order to arrive at a finite result in the WW approximation.
One therefore cannot pick out an individual twist-3 GPD and study its impact on observables or fit it to data, and at the same time use the WW approximation for the remaining twist-3 GPDs.
In such a case one would be left with an ill-defined framework.
This discussion holds for any of the twist-3 vector and axial-vector GPDs.
\\
\indent
One might wonder whether discontinuous twist-3 GPDs are an artifact of the WW approximation.
However, in the next section we will show that also in the QTM twist-3 GPDs are discontinuous, which suggests that such discontinuities are a general of these functions.
Speculations along those lines can be found in the literature already --- see for instance~\cite{Kivel:2000rb, Belitsky:2005qn}.

%%%%%%%%%%%%%%%%%%%%
\section{Twist-3 GPDs in the quark-target model}
\label{s:QTM}
%%%%%%%%%%%%%%%%%%%%
Twist-2 GPDs~\cite{Mukherjee:2002pq, Mukherjee:2002xi, Chakrabarti:2004ci, Meissner:2007rx, Ji:2015qla} and certain twist-3 GPDs~\cite{Mukherjee:2002pq, Mukherjee:2002xi} have been calculated previously in the QTM. 
Most of these studies have only considered the DGLAP region $x > \xi$.
Here we investigate for the first time if (twist-3) GPDs in the QTM are continuous at $x = \pm \, \xi$.
\\
\indent
We use the light-cone gauge $A^+ = 0$ and work to lowest nontrivial order in perturbation theory. 
We do not consider virtual graphs as they contribute for $x = 1$ only.
For the transverse part of the correlator in Eq.~\eqref{e:vector} one finds
\begin{equation}
F_\perp^\mu = - i \, \frac{C_F \, g^2}{(2 \pi)^4} \, P^+ \int_{- \infty}^{\infty} d k^- \, d^2 \vec{k}_\perp \, \frac{N_\perp^\mu}{D} \,,
\label{e:vector_QTM}
\end{equation}
with the numerator and denominator given by
\begin{eqnarray}
N_\perp^\mu & = & - \, \bar{u}(p') \, \gamma^\alpha \, \bigg( \slashed{k} + \frac{\slashed{\Delta}}{2} + m \bigg) \, \gamma_\perp^\mu \, \bigg( \slashed{k} - \frac{\slashed{\Delta}}{2} + m \bigg) \, \gamma^\beta \, u(p) \, D_{\alpha \beta}(P - k) \,,
\label{e:num_QTM_vector}
\\[0.2cm]
D & = & \bigg[ \Big(k - \frac{\Delta}{2} \Big)^2 - m^2 + i \varepsilon \bigg] \bigg[ \Big(k + \frac{\Delta}{2} \Big)^2 - m^2 + i \varepsilon \bigg] \big[(P - k)^2 + i \varepsilon \big] \,,
\label{e:den_QTM}
\end{eqnarray}
and the gluon polarization sum
\begin{equation}
D^{\mu\nu}(k) = - \, g^{\mu\nu} + \frac{k^\mu \, n^\nu + k^\nu \, n^\mu}{k \cdot n} \,.
\label{e:polarization_sum}
\end{equation}
In Eq.~\eqref{e:vector_QTM}, $g$ denotes the strong coupling constant (with $\alpha_s = \frac{g^2}{4 \pi}$), and $C_F = \frac{4}{3}$. 
To obtain the axial-vector correlator in Eq.~\eqref{e:axial} one has to replace $\gamma_\perp^\mu$ by $\gamma_\perp^\mu \gamma_5$ in~\eqref{e:num_QTM_vector}.
We denote the corresponding numerator by $\widetilde{N}_\perp^\mu$. 
\\
\indent
In this model calculation one encounters two types of $k^-$ integrals:
\begin{equation}
\big\{I; \, I^k \big \} = \int_{- \infty}^{\infty} dk^- \, \frac{\big\{1; \, k^- \big \}}{D}
= \frac{1}{C} \int_{- \infty}^{\infty} dk^- \, \frac{\big\{1; \, k^- \big \}}{(k^- - k_1^-) (k^- - k_2^-) (k^- - k_3^-)} \,,
\label{e:int_kminus}
\end{equation}
with the $k$ independent factor
\begin{equation}
C = - 8 \, (x + \xi) (x - \xi) (1 - x) (P^+)^3 \,,
\label{e:fact_C}
\end{equation}
and 
\begin{eqnarray}
k_1^- & = & \frac{\Delta^-}{2} + \frac{\big(\vec{k}_\perp - \frac{\vec{\Delta}_\perp}{2} \big)^2 + m^2 - i \varepsilon}{2 (x + \xi) P^+} \,,
\label{e:k1}
\\[0.2cm]
k_2^- & = & - \, \frac{\Delta^-}{2} + \frac{\big(\vec{k}_\perp + \frac{\vec{\Delta}_\perp}{2} \big)^2 + m^2 - i \varepsilon}{2 (x - \xi) P^+} \,,
\label{e:k2}
\\[0.2cm]
k_3^- & = & P^- - \frac{\vec{k}_\perp^2 - i \varepsilon}{2 (1 - x) P^+} \,.
\label{e:k3}
\end{eqnarray}
From Eqs.~\eqref{e:k1}--\eqref{e:k3} it is obvious that the position of the poles of the denominator in~\eqref{e:int_kminus} depends on the value of $x$.
We distinguish three regions for $x$, and evaluate the integrals in~\eqref{e:int_kminus} by using contour integration.
For the integral $I$ one readily obtains
\begin{equation}
I = 
\begin{cases}
I_1 = \frac{2 \pi i}{C} \frac{1}{(k_1^- - k_3^-) (k_2^- - k_3^-)} \,, \; \textrm{for} \;\, x > \xi \,, \\
I_2 = - \frac{2 \pi i}{C} \frac{1}{(k_1^- - k_2^-) (k_1^- - k_3^-)} \,, \; \textrm{for} \;\, - \xi \le x \le \xi \,, \\
I_3 = 0 \,, \; \textrm{for} \;\, x < - \, \xi \,.
\end{cases}
\label{e:I_res}
\end{equation}
According to~\eqref{e:I_res} the functional form of $I$ is different for the three regions.
However, $I$ is continuous at $x = \pm \, \xi$. 
To verify this statement for $x = + \, \xi$ one can consider the difference $I_1 - I_2$, which is given by
\begin{equation}
I_1 - I_2 = \frac{2 \pi i}{C} \frac{1}{(k_1^- - k_2^-) (k_2^- - k_3^-)} \,.
\label{e:I12_diff}
\end{equation}
Because of the two factors of $k_2^-$ in the denominator in~\eqref{e:I12_diff} that difference is proportional to $(x - \xi)$ and therefore vanishes for $x = + \, \xi$.
Likewise, $I_2$ in~\eqref{e:I_res} is proportional to $(x + \xi)$ due to the two factors of $k_1^-$ in the denominator, and it therefore vanishes at $x = - \, \xi$.
We also mention that the result $I = 0$ for $x < - \, \xi$ was expected since, at order ${\cal O}(g^2)$, there cannot be an antiquark distribution for a quark target.
\\
\indent
We now shift our attention to the integral $I^k$ in Eq.~\eqref{e:int_kminus} for which one finds
\begin{equation}
I^k = 
\begin{cases}
I_1^k = \frac{2 \pi i}{C} \frac{k_3^-}{(k_1^- - k_3^-) (k_2^- - k_3^-)} \,, \; \textrm{for} \;\, x > \xi \,, \\
I_2^k =  - \frac{2 \pi i}{C} \frac{k_1^-}{(k_1^- - k_2^-) (k_1^- - k_3^-)} \,, \; \textrm{for} \;\, - \xi \le x \le \xi \,, \\
I_3^k = 0 \,, \; \textrm{for} \;\, x < - \, \xi \,.
\end{cases}
\label{e:Ik_res}
\end{equation} 
It turns out that $I^k$ is discontinuous at $x = \pm \, \xi$.
In order to illustrate this point and to get a simple expression for the discontinuities we write
\begin{eqnarray}
I_1^k - I_2^k & = &
\frac{2 \pi i}{C} \frac{k_2^-}{(k_1^- - k_2^-) (k_2^- - k_3^-)}
= \frac{2 \pi i}{C} \bigg[ \frac{k_3^-}{(k_1^- - k_2^-) (k_2^- - k_3^-)} + \frac{1}{k_1^- - k_2^-} \bigg] \,,
\label{e:Ik12_diff}
\\[0.2cm]
I_2^k & = & - \, \frac{2 \pi i}{C} \bigg[ \frac{k_3^-}{(k_1^- - k_2^-) (k_1^- - k_3^-)} + \frac{1}{k_1^- - k_2^-} \bigg] \,.
\label{e:Ik_2}
\end{eqnarray}
The first term in the square brackets on the r.h.s.~of~\eqref{e:Ik12_diff} vanishes for $x = + \, \xi$, and the first term in the square brackets of~\eqref{e:Ik_2} vanishes for $x = - \, \xi$.
However, the expression $C (k_1^- - k_2^-)$ is finite at $x = \pm \, \xi$, leading to a discontinuous result for $I^k$ at these two kinematical points.
Therefore, GPDs in the QTM are generally discontinuous if they contain the integral $I^k$. 
It turns out that twist-2 GPDs in this model do contain $I^k$, but this integral is accompanied by the factor $(x^2 - \xi^2)$, and therefore no discontinuity occurs.
On the other hand, we show that most of the twist-3 vector and axial-vector GPDs are discontinuous.
We note in passing that in the QTM numerator terms proportional to $k^-$ can also lead to delta-function singularities at $x = 0$ for forward twist-3 parton distributions~\cite{Burkardt:1995ts, Burkardt:2001iy}.
\\
\indent
In the following we exclusively consider the $k^-$ dependent terms in the numerators $N_\perp^\mu$ and $\widetilde{N}_\perp^\mu$.
To find such terms for the various twist-3 GPDs, we rewrite the relevant contributions by using the Dirac bilinears in Eqs.~\eqref{e:tw3_vector} and~\eqref{e:tw3_axial} as basis vectors.
We skip the details of the calculation and merely mention that we have used
\begin{equation}
\int_{- \infty}^{\infty} d^2 \vec{k}_\perp \, \frac{k_\perp^\mu}{D}
= \Delta_\perp^\mu \int_{- \infty}^{\infty} d^2 \vec{k}_\perp \, \frac{\vec{k}_\perp \cdot \vec{\Delta}_\perp}{\vec{\Delta}_\perp^2 \, D} \,,
\label{e:kperp_int}
\end{equation}
which results from the fact that the integral on the l.h.s.~of~\eqref{e:kperp_int} must be proportional to $\Delta_\perp^\mu$.
Calculating the two numerators provides
\begin{eqnarray}
N_\perp^\mu & = & \frac{2 P^+ k^-}{1 - x} \, \bigg[ 
4 (1 - \xi^2) \, h_\perp^\mu 
- 2 (1 - 2x) \, \frac{\vec{k}_\perp \cdot \vec{\Delta}_\perp}{\vec{\Delta}_\perp^2} \, \Delta_\perp^\mu \frac{h^+}{P^+}
\nonumber \\[0.2cm]
&& - \, \bigg( 1 - x - 2\xi \, \frac{\vec{k}_\perp \cdot \vec{\Delta}_\perp}{\vec{\Delta}_\perp^2} \bigg) \tilde{\Delta}_\perp^\mu \frac{\tilde{h}^+}{P^+} \bigg] + \ldots \,,
\label{e:num_QTM_vector_res}
\\[0.2cm]
\widetilde{N}_\perp^\mu & = & \frac{2 P^+ k^-}{1 - x} \, \bigg[ 
4 x (1 - \xi^2) \, \tilde{h}_\perp^\mu 
+ 2 \bigg( \xi (1 - x) - (1 - 2x) \, \frac{\vec{k}_\perp \cdot \vec{\Delta}_\perp}{\vec{\Delta}_\perp^2} \bigg) \Delta_\perp^\mu \frac{\tilde{h}^+}{P^+}
\nonumber \\[0.2cm]
&& + \, \bigg( 1 - x + 2\xi \, \frac{\vec{k}_\perp \cdot \vec{\Delta}_\perp}{\vec{\Delta}_\perp^2} \bigg) \tilde{\Delta}_\perp^\mu \frac{h^+}{P^+} \bigg] + \ldots \,,
\label{e:num_QTM_axial_res}
\end{eqnarray}
where the dots in~\eqref{e:num_QTM_vector_res} and~\eqref{e:num_QTM_axial_res} indicate contributions without $k^-$ dependence.
(Higher powers of $k^-$ do not occur.)
Comparing the expressions in~\eqref{e:num_QTM_vector_res} and~\eqref{e:num_QTM_axial_res} with the parameterizations in~\eqref{e:tw3_vector} and~\eqref{e:tw3_axial}, respectively, one finds $k^-$ dependence for all twist-3 GPDs except $G_1$ and $\widetilde{G}_1$.
The above discussion about the integral $I^k$ therefore implies discontinuous twist-3 GPDs in the QTM.
This suggests that the discontinuities of twist-3 GPDs discussed in the previous section should not be considered an artifact of the WW approximation but rather a general feature of these functions.
That $G_1$ and $\widetilde{G}_1$ in the QTM at lowest order are continuous may be caused by the simplicity of the model.
\\
\indent
We now investigate if the results in the QTM are compatible with factorization for the amplitude in Eq.~\eqref{e:long_comb_res}.
The first linear combination of twist-3 GPDs in that equation is obviously continuous in the QTM.
Given that in the model calculation twist-2 GPDs and $\widetilde{G}_1$ are continuous, for the second linear combination of GPDs in Eq.~\eqref{e:long_comb_res} one just needs to consider
\begin{eqnarray}
A_2 & = & G_2 \, C^{+} - \frac{1}{\xi} \, \widetilde{G}_2 \, C^{-}
\nonumber \\[0.2cm]
& = & - i \, \frac{C_F \, g^2}{(2 \pi)^4} \, P^+ \int_{- \infty}^{\infty} d k^- \, d^2 \vec{k}_\perp \,
8 P^+ k^- \, \frac{1 - \xi^2}{1 - x} \, \bigg( C^{+} - \frac{x}{\xi} \, C^{-} \bigg) \frac{1}{D} + \ldots 
\nonumber \\[0.2cm]
& = & 0 + \ldots \,.
\label{e:A2}
\end{eqnarray}
The $k^-$ dependence in $A_2$ vanishes since $(\xi \, C^{+} - x \, C^{-}) = 0$.
Therefore the integration upon $x$ in Eq.~\eqref{e:long_comb_res} is well-defined for the linear combination $A_2$. 
For the third linear combination of GPDs in~\eqref{e:long_comb_res} one has
\begin{eqnarray}
A_3 & = & G_3 \, C^{+} - \widetilde{G}_4 \, C^{-}
\nonumber \\[0.2cm]
& = &  i \, \frac{C_F \, g^2}{(2 \pi)^4} \, P^+ \int_{- \infty}^{\infty} d k^- \, d^2 \vec{k}_\perp \,
\frac{2 P^+ k^-}{1 - x} \,
\bigg[ 2 (1 - 2x) \, \frac{\vec{k}_\perp \cdot \vec{\Delta}_\perp}{\vec{\Delta}_\perp^2} \, C^{+}
\nonumber \\[0.2cm]
& & + \, \bigg( 1 - x + 2\xi \, \frac{\vec{k}_\perp \cdot \vec{\Delta}_\perp}{\vec{\Delta}_\perp^2} \bigg) C^{-} \bigg]
\frac{1}{D} + \ldots 
\nonumber \\[0.2cm]
& = & \frac{1}{x - \xi + i \varepsilon} \, {\cal A}_{3, + \xi} + \frac{1}{x + \xi - i \varepsilon} \, {\cal A}_{3, - \xi} \,.
\label{e:A3}
\end{eqnarray}
The $x$ integration of $A_3$ can be performed provided that the function ${\cal A}_{3, + \xi}$ in~\eqref{e:A3} is continuous at $x = + \, \xi$ and ${\cal A}_{3, - \xi}$ is continuous at $x = - \, \xi$.
After carrying out the $k^-$ integral one obtains for the discontinuity of ${\cal A}_{3, + \xi}$ at $x = + \, \xi$:
\begin{eqnarray}
\lefteqn{\lim_{\delta \to 0} \big[ {\cal A}_{3, + \xi} (x = \xi + \delta) - {\cal A}_{3, + \xi} (x = \xi - \delta) \big]}
\nonumber \\[0.2cm]
&& = i \, \frac{C_F \, g^2}{(2 \pi)^4} \, 2 (P^+)^2 \, \int_{- \infty}^{\infty} d^2 \vec{k}_\perp \,
\bigg( 1+ 2 \, \frac{\vec{k}_\perp \cdot \vec{\Delta}_\perp}{\vec{\Delta}_\perp^2} \bigg) \, \big( I_1^k - I_2^k \big)\big|_{x = + \, \xi}
\nonumber \\[0.2cm]
&& = - \, \frac{C_F \, g^2}{(2 \pi)^3} \, \frac{1}{4 \xi (1 - \xi)} \, \frac{1}{\vec{\Delta}_\perp^2}
\int_{- \infty}^{\infty} d^2 \vec{k}_\perp \,
\frac{\vec{\Delta}_\perp^2 + 2 \, \vec{k}_\perp \cdot \vec{\Delta}_\perp}{\big(\vec{k}_\perp + \frac{\vec{\Delta}_\perp}{2} \big)^2 + m^2}
\nonumber \\[0.2cm]
&& = 0 \,.
\label{e:A3_xi_limit}
\end{eqnarray}
To derive the result in Eq.~\eqref{e:A3_xi_limit} we have used
\begin{equation}
\big( I_1^k - I_2^k \big)\big|_{x = + \, \xi} 
= \frac{2 \pi i}{C} \, \frac{1}{k_1^- - k_2^-} \Big|_{x = + \xi}
= \frac{\pi i}{ 4 \xi (1 - \xi) \, (P^+)^2 \big[ \big(\vec{k}_\perp + \frac{\vec{\Delta}_\perp}{2} \big)^2 + m^2 \big] } \,,
\label{e:Ik12_diff_xi}
\end{equation}
and that the integral upon the transverse momentum vanishes as can be shown by using the integration variable $\vec{l}_\perp = \vec{k}_\perp + \frac{\vec{\Delta}_\perp}{2}$.
With an analogous discussion one finds that ${\cal A}_{3, - \xi}$ is continuous at $x = - \, \xi$.
A very similar analysis shows that also the last linear combination of GPDs in Eq.~\eqref{e:long_comb_res} can be integrated upon $x$.
The results in the QTM are therefore compatible with factorization for DVCS at twist-3 accuracy, despite the discontinuous GPDs.
This finding further supports the hypothesis of factorization of the twist-3 DVCS amplitude.
In that regard our study is complimentary to the NLO calculation of DVCS in the WW approximation~\cite{Kivel:2003jt}.

%%%%%%%%%%%%%%%%%%%%
\section{Summary and discussion} 
\label{s:summary}
%%%%%%%%%%%%%%%%%%%%
At twist-3 accuracy, the amplitude for DVCS off the nucleon contains twist-2 as well as twist-3 GPDs.
Knowledge about twist-3 GPDs is therefore important for a reliable estimate of power corrections to the leading-twist DVCS amplitude.
Moreover, for a number of reasons, twist-3 GPDs are interesting in their own right~\cite{Penttinen:2000dg, Hatta:2012cs, Burkardt:2008ps, Burkardt:prep, Lorce:2014mxa, Bhoonah:2017olu, Rajan:2016tlg, Rajan:2017xlo}.
However, we have pointed out that in DVCS one cannot measure any individual twist-3 GPD.
This implies, in particular, that the kinetic OAM $L_{\rm kin}^q$ of quarks cannot be studied directly in DVCS via the twist-3 GPD $G_2$.
Accessible are only linear combinations involving both vector and axial-vector twist-3 GPDs.
We have made explicit these linear combinations.
\\
\indent
It has been known for quite some time that in the WW approximation twist-3 GPDs can exhibit discontinuities at $x = \pm \, \xi$~\cite{Kivel:2000rb}.
We have derived the WW approximation of the eight twist-3 vector and axial-vector GPDs of the nucleon.
All of them are discontinuous at both $x = + \, \xi$ and $x = - \, \xi$.
But the discontinuities cancel in the linear combinations of GPDs that enter the DVCS amplitude so that factorization is preserved.
\\
\indent
We have also computed the twist-3 GPDs in the QTM at lowest order in perturbation theory, and we have found discontinuities for most of these GPDs.
This result illustrates that these discontinuities are not artifacts of the WW approximation as the QTM (implicitly) includes both quark-gluon-quark correlations as well as quark mass terms, suggesting that discontinuities may be a more general feature of twist-3 GPDs. 
In the QTM, like for the WW approximation, the discontinuities cancel in the DVCS amplitude, which further supports the hypothesis of factorization at twist-3 accuracy.
\\
\indent
In the case of twist-2 GPDs it is known that QCD evolution does eliminate potential discontinuities (see, e.g., Ref.~\cite{Broniowski:2007si} for an explicit numerical demonstration).
Evolution equations for twist-3 GPDs do presently not exist. 
On the other hand, the splitting of a quark into a quark plus gluon is part of the QCD evolution, and our explicit perturbative calculation in the QTM has demonstrated that the splitting itself gives rise to discontinuities for individual twist-3 GPDs.
This suggests that discontinuities are generated by evolution rather than washed out, and, most likely, one can derive well-behaved evolution equations only for suitable linear combinations of twist-3 GPDs.
This interesting topic of course requires further investigation.
Moreover, we point out that also in the WW approximation the discontinuities exist at any scale because that approximation applies for any scale.
\\
\indent
Since only linear combinations of twist-3 GPDs can be accessed in DVCS, one may be tempted to estimate certain twist-3 GPDs in models and then fit other twist-3 GPDs of interest to DVCS data.
However, such an approach is questionable if not impossible:
If a model for twist-3 GPDs does not exhibit discontinuities it apparently misses an important feature of these functions.
On the other hand, if a model leads to discontinuous twist-3 GPDs, individual GPDs cannot be treated as arbitrary functions to be fitted to data.
\\
\indent
Our work suggests directions for further research.
For instance, one should try to explore the physics contained in the linear combinations of twist-3 GPDs that can be addressed in DVCS.
Moreover, it is important to search for other processes through which twist-3 GPDs can be studied --- in order to address different (linear combinations of) GPDs and/or to identify processes for which discontinuities of GPDs at $x = \pm \, \xi$ do not spoil factorization. 
It has been pointed out earlier that, in general, discontinuous GPDs do not cause a problem for double DVCS (lepto-production of a di-lepton pair)~\cite{Kivel:2000rb}.
However, the count rate for double DVCS is low~\cite{Guidal:2002kt}.
But the interesting physics contained in twist-3 GPDs warrants further studies whose final goal is the measurement of these functions.

%%%%%%%%%%%%%%%%%%%%
\begin{acknowledgments}
This work has been supported by the Department of Energy under grant number DE-FG03-95ER40965 (F.A.~and M.B.), the National Science Foundation under grant number PHY-1516088 (A.M.), and the European Research Council (ERC) under the European Union's Horizon 2020 Research and Innovation Programme (grant agreement number 647981, 3DSPIN) (B.P.).
The work of M.B.~and A.M.~has also been supported by the U.S. Department of Energy, Office of Science, Office of Nuclear Physics, within the framework of the TMD Topical Collaboration.
B.P.~is grateful for the hospitality to the Centre de Physique Th\'eorique, \'Ecole Polytechnique, where part of this work was carried out, and acknowledges the support of the PICS program (Contract N. PICS07561) during her stay.
\end{acknowledgments}

%%%%%%%%%%%%%%%%%%%%
\appendix

%%%%%%%%%%%%%%%%%%%%
\section{Relations between Dirac bilinears} 
%%%%%%%%%%%%%%%%%%%%
Here we list several relations between Dirac bilinears, which are all based on the Dirac equation --- see also, e.g., Refs.~\cite{Belitsky:2000vx, Belitsky:2005qn, Lorce:2017isp}.
The relations that we have used in this work are
\begin{eqnarray} 
i \varepsilon_{\perp \alpha}^\mu \, \tilde{h}_\perp^\alpha & = &
\frac{1}{\xi} \, h_\perp^\mu 
- \frac{1}{2 \xi} \,\tilde{\Delta}_\perp^\mu \frac{\tilde{h}^+}{P^+} \,,
\label{e:dirac_1a}
\\[0.2cm]
i \varepsilon_{\perp \alpha}^\mu \, h_\perp^\alpha & = &
\xi \, \tilde{h}_\perp^\mu 
+ \frac{1}{2} \,\Delta_\perp^\mu \frac{\tilde{h}^+}{P^+} \,,
\label{e:dirac_1b}
\\[0.2cm] 
\tilde{\Delta}_\perp^\mu \, \tilde{b} & = &
- \, \Delta_\perp^\mu \, b
+ \frac{\Delta_\perp^2}{2 \xi m} \, h_\perp^\mu
+ \frac{\bar{m}^2}{m} \, \Delta_\perp^\mu \frac{h^+}{P^+}
- \frac{t}{4 \xi m} \,\tilde{\Delta}_\perp^\mu \frac{\tilde{h}^+}{P^+} \,,
\label{e:dirac_2a}
\\[0.2cm]
\tilde{\Delta}_\perp^\mu \, b & = &
- \, \Delta_\perp^\mu \, \tilde{b}
+ \frac{\Delta_\perp^2}{2 m} \, \tilde{h}_\perp^\mu
+ \frac{\xi \bar{m}^2}{m} \, \Delta_\perp^\mu \frac{\tilde{h}^+}{P^+}
+ \frac{\bar{m}^2}{m} \, \tilde{\Delta}_\perp^\mu \frac{h^+}{P^+} \,,
\label{e:dirac_2b}
\\[0.2cm]
h^{\mu} & = & \frac{P^{\mu}}{m} \, b + e^{\mu} \,,
\label{e:dirac_3}
\\[0.2cm]
t^{+ \mu} & = & 
\frac{P^+}{2 \xi m} \bigg[ - 2 (1 - \xi^2) \, h_\perp^\mu + \xi \, \Delta_\perp^\mu \frac{h^+}{P^+} + \tilde{\Delta}_\perp^\mu \frac{\tilde{h}^+}{P^+} \bigg] \,,
\label{e:dirac_4a}
\\[0.2cm]
i \varepsilon_{\perp \alpha}^\mu \, t^{+ \alpha} & = & 
\frac{P^+}{2 m} \bigg[ - 2 (1 - \xi^2) \, \tilde{h}_\perp^\mu + \xi \, \Delta_\perp^\mu \frac{\tilde{h}^+}{P^+} + \tilde{\Delta}_\perp^\mu \frac{h^+}{P^+} \bigg] \,.
\label{e:dirac_4b}
\end{eqnarray}

%%%%%%%%%%%%%%%%%%%%
\section{Comparing different conventions for twist-3 GPDs}
%%%%%%%%%%%%%%%%%%%%
We compare here the notation for twist-3 GPDs from Ref.~\cite{Kiptily:2002nx}, which we have used in the main body of this paper, with the notation of 
Refs.~\cite{Belitsky:2001yp, Belitsky:2001ns} and of Ref.~\cite{Meissner:2009ww}.
In Refs.~\cite{Belitsky:2001yp, Belitsky:2001ns} the correlators $F^\mu$ and $\widetilde{F}^\mu$ in Eqs.~\eqref{e:vector} and~\eqref{e:axial} are parameterized according to\footnote{In Ref.~\cite{Belitsky:2005qn} the same GPD notation is used, but with the momentum transfer defined as $\Delta' = p - p' = - \Delta$.}
\begin{eqnarray}
F^\mu & = & P^\mu \frac{h^+}{P^+} \, H + P^\mu \frac{e^+}{P^+} \,E
\nonumber \\[0.2cm]
&& + \, \Delta_\perp^\mu \frac{h^+}{2P^+}  \, H_{+}^3 + \Delta_\perp^\mu \frac{e^+}{2P^+} \, E_{+}^3
+ \tilde{\Delta}_\perp^\mu \frac{\tilde{h}^+}{2P^+} \, \widetilde{H}_{-}^3 + \tilde{\Delta}_\perp^\mu \frac{\tilde{e}^+}{2P^+} \, \widetilde{E}_{-}^3 \,,
\label{e:tw3_vector_bm}
\\[0.2cm]
\widetilde{F}^\mu & = & P^\mu \frac{\tilde h^+}{P^+} \, \widetilde{H} + P^\mu \frac{\tilde{e}^+}{P^+} \, \widetilde{E}
\nonumber \\[0.2cm]
&& + \, \Delta_\perp^\mu \frac{\tilde{h}^+}{2P^+} \, \widetilde{H}_{+}^3 + \Delta_\perp^\mu \frac{\tilde{e}^+}{2P^+} \, \widetilde{E}_{+}^3
+ \tilde{\Delta}_\perp^\mu \frac{h^+}{2P^+} \, H_{-}^3 +  \tilde{\Delta}_\perp^\mu \frac{e^+}{2P^+} \, E_{-}^3 \,.
\label{e:tw3_axial_bm}
\end{eqnarray}
In order to relate the twist-3 GPDs in~\eqref{e:tw3_vector_bm} and~\eqref{e:tw3_axial_bm} to the ones in Eqs.~\eqref{e:tw3_vector} and~\eqref{e:tw3_axial} we use $\tilde{e}^+ = \Delta^+ \tilde{b} / (2m)$ (see~\eqref{e:bilinears}), the relation~\eqref{e:dirac_3} for $\mu = +$, as well as Eqs.~\eqref{e:dirac_2a} and~\eqref{e:dirac_2b}.
One finds
\begin{equation} 
\begin{aligned}
H_{+}^3 & = G_1 + \frac{\xi t}{\Delta_\perp^2} \, (H + E + G_2) + 2 \, G_3 \,,  \qquad &
E_{+}^3 & = - \, G_1 - \frac{4 \xi m^2}{\Delta_\perp^2} \, (H + E + G_2) \,,
\\[0.2cm]
\widetilde{H}_{-}^3 & = \frac{t}{\Delta_\perp^2} \, (H + E + G_2) + 2 \, G_4 \,, &
\widetilde{E}_{-}^3 & = - \frac{4 m^2}{\Delta_\perp^2} \, (H + E + G_2) \,,
\\[0.2cm]
\widetilde{H}_{+}^3 & = - \, \frac{4 \xi \bar{m}^2}{\Delta_\perp^2} \, (\widetilde{H} + \widetilde{G}_2) + 2 \, \widetilde{G}_3 \,, &
\widetilde{E}_{+}^3 & = - \, \frac{1}{\xi} \, ( \widetilde{E} + \widetilde{G}_1) - \frac{4 m^2}{\xi \Delta_\perp^2} \, (\widetilde{H} + \widetilde{G}_2) \,,
\\[0.2cm]
H_{-}^3 & = \frac{t}{\Delta_\perp^2} \, ( \widetilde{H} + \widetilde{G}_2 ) + 2 \, \widetilde{G}_4 \,, &
E_{-}^3 & = - \, \frac{4 m^2}{\Delta_\perp^2} \, ( \widetilde{H} + \widetilde{G}_2) \,.
\end{aligned}
\label{e:relation_bm}
\end{equation}
The inversion of the set of equations in~\eqref{e:relation_bm} reads
\begin{equation} 
\begin{aligned}
G_1 & = - \, E_{+}^3 + \xi \widetilde{E}_{-}^3 \,, &
G_2 & = - \, (H + E) - \frac{\Delta_\perp^2}{4 m^2} \, \widetilde{E}_{-}^3 \,,
\\[0.2cm]
G_3 & = \frac{1}{2} \, \bigg( H_{+}^3 + E_{+}^3 - \frac{\xi \bar{m}^2}{m^2} \, \widetilde{E}_{-}^3 \bigg) \,,  \qquad &
G_4 & = \frac{1}{2} \, \bigg( \widetilde{H}_{-}^3 + \frac{t}{4m^2} \, \widetilde{E}_{-}^3 \bigg) \,,
\\[0.2cm]
\widetilde{G}_1 & = - \, \widetilde{E} - \xi \widetilde{E}_{+}^3 + E_{-}^3 \,, &
\widetilde{G}_2 & = - \, \widetilde{H} - \frac{\Delta_\perp^2}{4 m^2} \, E_{-}^3  \,,
\\[0.2cm]
\widetilde{G}_3 & = \frac{1}{2} \, \bigg( \widetilde{H}_{+}^3 - \frac{\xi \bar{m}^2}{m^2} \, E_{-}^3 \bigg) \,, &
\widetilde{G}_4 & = \frac{1}{2} \, \bigg( H_{-}^3 + \frac{t}{4 m^2} \, E_{-}^3 \bigg) \,.
\end{aligned}
\label{e:relation_bm_inv}
\end{equation}

In Ref.~\cite{Meissner:2009ww} both chiral-even and chiral-odd twist-3 GPDs have been defined, where the chiral-even ones are given by \begin{eqnarray}
F^\mu & = & P^\mu \frac{h^+}{P^+} \, H + P^\mu \frac{e^+}{P^+} \, E 
\nonumber \\[0.2cm]
&& + \, \frac{m}{P^+} \bigg[ t^{+\mu} \, H_{2T} + \frac{1}{2m} ( \Delta_\perp^\mu h^+ - \Delta^+ h_\perp^\mu ) \, E_{2T} + \frac{P^+}{m^2} \,\Delta_\perp^\mu b \, \widetilde{H}_{2T} - \frac{P^+}{m} \, h_\perp^\mu \, \widetilde{E}_{2T} \bigg] \,,
\label{e:tw3_vector_mms}
\\[0.2cm]
\widetilde{F}^\mu & = & P^\mu \frac{\tilde{h}^+}{P^+} \, \widetilde{H} + P^\mu \frac{\tilde{e}^+}{P^+} \, \widetilde{E} 
\nonumber \\[0.2cm]
&& - \, i \varepsilon_{\perp \alpha}^\mu \, \frac{m}{P^+} \bigg[ t^{+ \alpha} \, H_{2T}' + \frac{1}{2m} ( \Delta_\perp^\alpha h^+ - \Delta^+ h_\perp^\alpha ) \, E_{2T}' + \frac{P^+}{m^2} \,\Delta_\perp^\alpha b \, \widetilde{H}_{2T}' - \frac{P^+}{m} \, h_\perp^\alpha \, \widetilde{E}_{2T}' \bigg] \,, \qquad
\label{e:tw3_axial_mms}
\end{eqnarray}
with $t^{\mu\nu}$ defined in the paragraph after~\eqref{e:bilinears}.
Using Eqs.~\eqref{e:dirac_1b},~\eqref{e:dirac_2b},~\eqref{e:dirac_4a},~\eqref{e:dirac_4b} one finds
\begin{equation} 
\begin{aligned}
H_{2T} & = 2 \, \xi G_4 \,, \quad &
E_{2T} & = 2 \, (G_3 -\xi G_4) \,,
\\[0.2cm]
\widetilde H_{2T} & = \frac{1}{2} \, G_1 \,, &
\widetilde E_{2T} & = \, - (H + E + G_2) + 2 \, (\xi G_3 - G_4) \,,
\\[0.2cm]
H_{2T}' & =  \frac{t}{4 m^2} \, (\widetilde{E} + \widetilde{G}_1) + (\widetilde{H} + \widetilde{G}_2) & \qquad
E_{2T}' & = - \, (\widetilde{E} + \widetilde{G}_1) - (\widetilde{H} + \widetilde{G}_2) 
\\[0.2cm]
& \hspace{0.5cm} - 2 \, \xi \widetilde{G}_3 \,, & & \hspace{0.5cm} + 2 \, (\xi \widetilde{G}_3 - \widetilde{G}_4) \,,
\\[0.2cm]
\widetilde{H}_{2T}' & = \frac{1}{2} \, (\widetilde{E} + \widetilde{G}_1) \,, &
\widetilde{E}_{2T}' & = 2 \, (\widetilde{G}_3 - \xi \widetilde{G}_4) \,.
\end{aligned}
\label{e:relation_mms}
\end{equation}
The inversion of the set of equations in~\eqref{e:relation_mms} reads
\begin{equation} 
\begin{aligned}
G_1 & = 2 \, \widetilde{H}_{2T} \,, &
G_2 & = - \, (H + E) - \frac{1}{\xi}(1 - \xi^2) \, H_{2T} 
\\[0.2cm]
& & & \hspace{0.5cm} + \xi E_{2T} - \widetilde{E}_{2T}  \,,
\\[0.2cm]
G_3 & = \frac{1}{2} \, (H_{2T} + E_{2T}) \,, \quad &
G_4 & = \frac{1}{2\xi} \, H_{2T} \,,
\\[0.2cm]
\widetilde{G}_1 & = - \, \widetilde{E} + 2 \, \widetilde{H}_{2T}' \,, &
\widetilde{G}_2 & = - \, \widetilde{H} + (1 - \xi^2) \, H_{2T}' - \xi^2 \, E_{2T}' 
\\[0.2cm]
& & & \hspace{0.5cm} - \frac{\Delta_\perp^2}{2 m^2} \, \widetilde{H}_{2T}' + \xi \widetilde{E}_{2T}'  \,,
\\[0.2cm]
\widetilde{G}_3 & = - \, \frac{\xi}{2} \, ( H_{2T}' + E_{2T}' ) - \frac{\xi \bar{m}^2}{m^2} \, \widetilde{H}_{2T}'  & \qquad
\widetilde{G}_4 & = - \, \frac{1}{2} \, ( H_{2T}' + E_{2T}' ) - \frac{\bar{m}^2}{m^2} \, \widetilde{H}_{2T}'  \,.
\\[0.2cm]
& \hspace{0.5cm} + \frac{1}{2} \, \widetilde{E}_{2T}'  \,,  & & 
\end{aligned}
\label{e:relation_mms_inv}
\end{equation}

%%%%%%%%%%%%%%%%%%%%
\section{Twist-3 GPDs in the WW approximation}
%%%%%%%%%%%%%%%%%%%%
Here we present to WW term for the twist-3 vector and axial-vector GPDs of a spin-$\frac{1}{2}$ target.
Making use of the Dirac bilinears in~\eqref{e:bilinears}, the WW parts of the correlators in Eqs.~\eqref{e:vector} and~\eqref{e:axial}, which are given by twist-2 GPDs, take the form~\cite{Kivel:2000cn}
\begin{eqnarray}
F_{WW}^\mu(x,\xi,\Delta) & = &
\frac{1}{\xi} \, \Delta^\mu \frac{b}{2m} \, E(x,\xi) - \frac{1}{2\xi} \, \Delta^\mu \frac{h^+}{P^+} \, (H + E)(x,\xi)
\nonumber \\[0.2cm]
&& + \, \int_{-1}^{1}du \, G^\mu(u,\xi,\Delta) \, W_{+}(x,u,\xi)
+ i \varepsilon_{\perp \alpha}^\mu \int_{-1}^{1} du \, \widetilde{G}^\alpha (u,\xi,\Delta) \, W_{-}(x,u,\xi) \, ,
\label{e:vector_ww}
\\[0.2cm] 
\widetilde{F}_{WW}^\mu(x,\xi,\Delta) & = &
\Delta^\mu \frac{\tilde{b}}{2 m} \, \widetilde{E}(x,\xi) - \frac{1}{2\xi} \, \Delta^\mu \frac{\tilde h^+}{P^+} \, \widetilde{H}(x,\xi)
\nonumber \\[0.2cm] 
&& + \int_{-1}^{1}du \, \widetilde{G}^\mu(u,\xi,\Delta) \, W_{+}(x,u,\xi)
+ i \varepsilon_{\perp \alpha}^\mu \int_{-1}^{1} du \, G^\alpha (u,\xi,\Delta) \, W_{-}(x,u,\xi) \,, \qquad 
\label{e:axial_ww}
\end{eqnarray}
with
\begin{eqnarray}
G^\mu(u,\xi,\Delta) & = & 
h_\perp^\mu \, (H + E)(u,\xi) 
+ \frac{1}{\xi} \, \Delta_\perp^\mu \frac{b}{2m} \, {\cal D}_{u,\xi} \big[ E(u,\xi) \big]
\nonumber \\[0.2cm]
&& - \, \frac{1}{2 \xi} \, \Delta_\perp^\mu \frac{h^+}{P^+} \, {\cal D}_{u,\xi} \big[ (H + E) (u,\xi) \big] \,,
\label{e:G_ww}
\\[0.2cm]
\widetilde{G}^\mu(u,\xi,\Delta) & = & 
\tilde{h}_\perp^\mu \, \widetilde{H}(u,\xi) 
+ \frac{1}{\xi} \, \Delta_\perp^\mu \frac{\tilde{b}}{2m} \, {\cal D}_{u,\xi} \big[ \xi \widetilde{E}(u,\xi) \big]
- \frac{1}{2 \xi} \, \Delta_\perp^\mu \frac{\tilde{h}^+}{P^+} \, {\cal D}_{u,\xi} \big[ \widetilde{H}(u,\xi) \big] \,,
\label{e:tG_ww}
\end{eqnarray}
and the differential operator
\begin{equation}
{\cal D}_{u,\xi} = u \frac{\partial}{\partial u} + \xi \frac{\partial}{\partial \xi} \,.
\label{e:diff_op}
\end{equation}
The so-called WW kernels $W_{\pm}(x,u,\xi)$ in Eqs.~\eqref{e:vector_ww} and~\eqref{e:axial_ww} are defined as~\cite{Kivel:2000cn}
\begin{eqnarray}
W_{\pm}(x,u,\xi) & = & 
\frac{1}{2 (u - \xi)} \Big[ \theta(x > \xi) \, \theta(u > x) - \theta(x < \xi) \, \theta(u < x)  \Big] 
\nonumber \\[0.2cm]
&& \pm \, \frac{1}{2 (u + \xi)} \Big[ \theta(x > - \, \xi) \, \theta(u > x)  - \theta(x < - \, \xi) \, \theta(u < x) \Big] \,.
\label{e:Wpm}
\end{eqnarray}
By means of the Eqs.~\eqref{e:dirac_1a} and~\eqref{e:dirac_2a} one can rewrite the expression in~\eqref{e:vector_ww} in terms of the Dirac bilinears used in the GPD decomposition of Eq.~\eqref{e:tw3_vector}.
This provides the WW approximation for the twist-3 vector GPDs:
\begin{eqnarray} 
G_1^{{\rm WW}}(x,\xi) & = &
\frac{1}{\xi} \, E(x,\xi) 
+ \frac{1}{\xi} \int_{-1}^{1} du \, W_{+}(x,u,\xi) \, {\cal D}_{u,\xi} \big[ E(u,\xi) \big]
\nonumber \\[0.2cm]
&& - \, \frac{1}{\xi} \int_{-1}^{1} du \, W_{-}(x,u,\xi) \, {\cal D}_{u,\xi} \big[ \xi \widetilde{E}(u,\xi) \big] \,,
\label{e:g1_ww}
\\[0.2cm] 
G_2^{{\rm WW}}(x,\xi) & = &
- \, (H + E)(x,\xi)
+ \int_{-1}^{1} du \, W_{+}(x,u,\xi) \, (H + E)(u,\xi)
\nonumber \\[0.2cm]
&& + \, \frac{1}{\xi^2} \int_{-1}^{1} du \, W_{-}(x,u,\xi) \, \bigg( \xi \widetilde{H}(u,\xi) + \frac{\Delta_\perp^2}{4 m^2} \, {\cal D}_{u,\xi} \big[ \xi \widetilde{E}(u,\xi) \big] \bigg) \,,
\label{e:g2_ww}
\\[0.2cm] 
G_3^{{\rm WW}}(x,\xi) & = &
- \, \frac{1}{2\xi} \, (H + E)(x,\xi)
- \frac{1}{2\xi} \int_{-1}^{1} du \, W_{+}(x,u,\xi) \, {\cal D}_{u,\xi} \big[ (H + E)(u,\xi) \big]
\nonumber \\[0.2cm]
&& + \, \frac{\bar{m}^2}{2 \xi m^2} \int_{-1}^{1} du \, W_{-}(x,u,\xi) \, {\cal D}_{u,\xi} \big[ \xi \widetilde{E}(u,\xi) \big] \,,
\label{e:g3_ww}
\\[0.2cm] 
G_4^{{\rm WW}}(x,\xi) & = &
-  \, \frac{1}{2 \xi^2} \int_{-1}^{1} du \, W_{-}(x,u,\xi) \, \bigg( {\cal D}_{u,\xi} \big[ \xi \widetilde{H}(u,\xi) \big]  
+ \frac{t}{4 m^2} \, {\cal D}_{u,\xi} \big[ \xi \widetilde{E}(u,\xi) \big] \bigg) \,. \quad
\label{e:g4_ww}
\end{eqnarray}
Likewise, by using the Eqs.~\eqref{e:dirac_1b} and~\eqref{e:dirac_2b} one can rewrite the expression in~\eqref{e:axial_ww} in terms of the Dirac bilinears that appear in the GPD decomposition of Eq.~\eqref{e:tw3_axial}.
This provides the WW approximation for the twist-3 axial-vector GPDs:
\begin{eqnarray} 
\widetilde{G}_1^{{\rm WW}}(x,\xi) & = &
\frac{1}{\xi} \int_{-1}^{1} du \, W_{+}(x,u,\xi) \, {\cal D}_{u,\xi} \big[ \xi \widetilde{E}(u,\xi) \big]
- \frac{1}{\xi} \int_{-1}^{1} du \, W_{-}(x,u,\xi) \, {\cal D}_{u,\xi} \big[ E(u,\xi) \big] \,, \qquad \;
\label{e:tg1_ww}
\\[0.2cm] 
\widetilde{G}_2^{{\rm WW}}(x,\xi) & = &
- \, \widetilde{H}(x,\xi)
+ \int_{-1}^{1} du \, W_{+}(x,u,\xi) \, \widetilde{H}(u,\xi)
\nonumber \\[0.2cm]
&& + \, \frac{1}{\xi} \int_{-1}^{1} du \, W_{-}(x,u,\xi) \, \bigg( \xi^2 (H + E)(u,\xi) + \frac{\Delta_\perp^2}{4 m^2} \, {\cal D}_{u,\xi} \big[ E(u,\xi) \big] \bigg) \,, \qquad \;
\label{e:tg2_ww}
\\[0.2cm] 
\widetilde{G}_3^{{\rm WW}}(x,\xi) & = &
- \, \frac{1}{2\xi} \, \widetilde{H}(x,\xi)
- \frac{1}{2\xi} \int_{-1}^{1} du \, W_{+}(x,u,\xi) \, {\cal D}_{u,\xi} \big[ \widetilde{H} (u,\xi) \big]
\nonumber \\[0.2cm]
&& + \, \frac{1}{2} \int_{-1}^{1} du \, W_{-}(x,u,\xi) \, \bigg( (H + E)(u,\xi) + \frac{\bar{m}^2}{m^2} \, {\cal D}_{u,\xi} \big[ E(u,\xi) \big] \bigg) \,,
\label{e:tg3_ww}
\\[0.2cm] 
\widetilde{G}_4^{{\rm WW}}(x,\xi) & = &
- \, \frac{1}{2 \xi} \int_{-1}^{1} du \, W_{-}(x,u,\xi) \, \bigg( {\cal D}_{u,\xi} \big[ H(u,\xi) \big] 
+ \frac{t}{4 m^2} \, {\cal D}_{u,\xi} \big[ E(u,\xi) \big] \bigg) \,.
\label{e:tg4_ww}
\end{eqnarray}
To the best of our knowledge, we have obtained for the first time a complete list of the WW terms for all twist-3 vector and axial-vector GPDs of the nucleon.
Based on the discussion in Sec.~\ref{s:WWapproximation} and the results in~\eqref{e:g1_ww}--\eqref{e:tg4_ww} one finds that all eight twist-3 GPDs are discontinuous at both $x = + \, \xi$ and $x = - \, \xi$.

The expressions in Eqs.~\eqref{e:g1_ww}--\eqref{e:tg4_ww} allow one to find the WW approximation for the four linear combinations of GPDs that appear on the r.h.s.~of Eq.~\eqref{e:long_comb_res}.
In Section~\ref{s:WWapproximation}, we have discussed the result for the first such linear combination.
For the other three cases one has
\begin{eqnarray} 
\lefteqn{\int_{-1}^{1} dx  \,  \bigg( (H + E + G_2^{{\rm WW}}) \, C^+ - \frac{\Delta_\perp^2}{4 \xi m^2} \, (\widetilde{E} + \widetilde{G}_1^{{\rm WW}}) \, C^- - \frac{1}{\xi} \, (\widetilde{H} + \widetilde{G}_2^{{\rm WW}}) \, C^-  \bigg)}
\nonumber \\[0.2cm]
& = & - \, \frac{\Delta_\perp^2}{4 \xi m^2} \int_{-1}^{1} dx \, \Big[ \widetilde{E}(x,\xi) \, C^-(x,\xi) \Big] 
\nonumber \\[0.2cm]
& + & \int_{-1}^{1} dx \int_{-1}^{1} du \, \Big[ W_{+}(x,u,\xi) \, C^+(x,\xi) - W_{-}(x,u,\xi) \, C^-(x,\xi) \Big] \, (H + E)(u,\xi)
\nonumber \\[0.2cm]
& + & \frac{1}{\xi^2} \int_{-1}^{1} dx \int_{-1}^{1} du \, \Big[ W_{-}(x,u,\xi) \, C^+(x,\xi) - W_{+}(x,u,\xi) \, C^-(x,\xi) \Big] 
\nonumber \\[0.2cm]
&& \hspace{2.5cm} \times \, \Big[  \xi \widetilde{H}(u,\xi) + \frac{\Delta_\perp^2}{4 m^2} \, {\cal D}_{u,\xi} \big[ \xi \widetilde{E} (u,\xi) \big]  \Big] \,,
\label{e:conv2}
\\[0.6cm]
\lefteqn{\int_{-1}^{1} dx  \,  \bigg( G_3^{{\rm WW}} \, C^+ - \frac{\bar{m}^2}{2 m^2} \, (\widetilde{E} + \widetilde{G}_1^{{\rm WW}}) \, C^- - \widetilde{G}_4^{{\rm WW}} \, C^- \bigg)}
\nonumber \\[0.2cm]
& = & - \frac{1}{2 \xi} \int_{-1}^{1} dx \, \Big[ (H + E)(x,\xi) \, C^+(x,\xi) 
+ \frac{\xi \bar{m}^2}{m^2} \, \widetilde{E}(x,\xi) \, C^-(x,\xi) \Big] 
\nonumber \\[0.2cm]
& - & \frac{1}{2\xi} \int_{-1}^{1} dx \int_{-1}^{1} du \, \Big[ W_{+}(x,u,\xi) \, C^+(x,\xi) - W_{-}(x,u,\xi) \, C^-(x,\xi) \Big] \, {\cal D}_{u,\xi} \big[ (H + E)(u,\xi) \big]
\nonumber \\[0.2cm]
& + & \frac{\bar{m}^2}{2 \xi m^2} \int_{-1}^{1} dx \int_{-1}^{1} du \, \Big[ W_{-}(x,u,\xi) \, C^+(x,\xi) - W_{+}(x,u,\xi) \, C^-(x,\xi) \Big] \, {\cal D}_{u,\xi} \big[ \xi \widetilde{E}(u,\xi) \big] \,, \qquad
\label{e:conv3}
\\[0.6cm]
\lefteqn{\int_{-1}^{1} dx  \, \bigg( G_4^{{\rm WW}} \, C^+ + \frac{t}{8 \xi m^2} \, (\widetilde{E} + \widetilde{G}_1^{{\rm WW}}) \, C^-  + \frac{1}{2\xi} \,(\widetilde{H} + \widetilde{G}_2^{{\rm WW}}) \, C^- - \widetilde{G}_3^{{\rm WW}} \, C^- \bigg)} 
\nonumber \\[0.2cm]
& = & \frac{1}{2 \xi} \int_{-1}^{1} dx \, \Big[ \widetilde{H}(x,\xi) \, C^-(x,\xi) + \frac{t}{4 m^2} \, \widetilde{E}(x,\xi) \, C^-(x,\xi) \Big] 
\nonumber \\[0.2cm]
& - & \frac{1}{2\xi^2} \int_{-1}^{1} dx \int_{-1}^{1} du \, \Big[ W_{-}(x,u,\xi) \, C^+(x,\xi) - W_{+}(x,u,\xi) \, C^-(x,\xi) \Big] 
\nonumber \\[0.2cm]
&& \hspace{2.5cm} \times \, \Big[  {\cal D}_{u,\xi} \big[ \xi \widetilde{H}(u,\xi) \big] + \frac{t}{4 m^2} \, {\cal D}_{u,\xi} \big[ \xi \widetilde{E}(u,\xi) \big] \Big] \,. \phantom{\int}
\label{e:conv4}
\end{eqnarray}
Like for Eq.~\eqref{e:conv1}, in the linear combinations in~\eqref{e:conv2}--\eqref{e:conv4} the WW kernels $W_{\pm}$ exclusively enter via the well-behaved integrals in~\eqref{e:conv_WC}.

%%%%%%%%%%%%%%%%%%%%

\end{document}